\documentstyle[aps,epsf]{revtex}
\newcommand{\be}{\begin{eqnarray}}
\newcommand{\ee}{\end{eqnarray}}

\draft
\begin{document}
\wideabs{
\title{New black holes in the brane-world?}
\author{Roberto Casadio$^{a}$, Alessandro Fabbri$^{b}$
and Lorenzo Mazzacurati$^{c}$}
\address{~}
\address{Dipartimento di Fisica, Universit\`a di
Bologna and I.N.F.N., Sezione di Bologna\\
via Irnerio 46, 40126 Bologna, Italy}
\maketitle
\begin{abstract}
It is known that the Einstein field equations in five dimensions
admit more general spherically symmetric black holes on the brane
than four-dimensional general relativity.
We propose two families of analytic solutions (with
$g_{tt}\not=-g_{rr}^{-1}$), parameterized by
the ADM mass and the PPN parameter $\beta$, which reduce
to Schwarzschild for $\beta=1$.
Agreement with observations requires $|\beta-1|\sim|\eta|\ll 1$.
The sign of $\eta$ plays a key role in the global
causal structure, separating metrics which behave like
Schwarzschild ($\eta<0$) from those similar to
Reissner-Nordstr\"om ($\eta>0$).
In the latter case, we find a family of black hole space-times
completely regular. 
\end{abstract}
\pacs{PACS: 04.70.-s, 04.70.Bw, 04.50.+h}
}
In recent years there has been a renewed interest in models
with extra dimensions in which the standard model fields
are confined to our four-dimensional world viewed as a
(infinitely thin) hypersurface (the brane) embedded in the
higher-dimensional space-time (the bulk) where (only) gravity
can propagate.
Of particular interest are cases where the extra dimensions
are infinitely extended but ``warped'' by the presence of a
non-vanishing bulk cosmological constant $\Lambda$ related to
the (singular) vacuum energy density of the brane
\cite{RS,kaloper} by the standard junction equations
\cite{israel}.
\par
In $D+1$ space-time dimensions a vacuum solution must satisfy
($\mu,\nu=0,\ldots,D$)
\be
{\mathcal R}_{\mu\nu}=\Lambda\,g_{\mu\nu}
\ .
\label{D+1eq}
\ee
On projecting the above equation on a time-like manifold of
codimension one (the brane) and introducing Gaussian normal
coordinates $x^i$ ($i=0,\ldots,D-1$) and $z$ ($z=0$
on the brane), one obtains the constraints (at $z=0$)
\be
{\mathcal R}_{iz}=0\ ,\ \ \
R=\lambda
\ ,
\label{Deq}
\ee
where $R$ is the $D$-dimensional Ricci scalar, $\lambda$ the
cosmological constant on the brane (we shall set $\lambda=0$
from now on, equivalently to the fine tuning
between $\Lambda$ and the brane tension \cite{RS}) and use has been
made of the necessary junction equations \cite{israel}.
For static solutions, one can view Eqs.~(\ref{Deq}) as the analogs
of the momentum and Hamiltonian constraints
in the ADM decomposition of the metric and their role is
therefore to select out admissible field configurations along
hypersurfaces of constant $z$.
Such field configurations will then be ``propagated'' off-brane
by the remaining Einstein Eqs.~(\ref{D+1eq}).
It is clear that the above ``Hamiltonian'' constraint is a
weaker requirement than the purely $D$-dimensional vacuum
equations $R_{ij}=0$ and, in fact, it is equivalent to
$R_{ij}=E_{ij}$ where $E_{ij}$ is (proportional to) the
(traceless) projection of the $D+1$-dimensional Weyl tensor
on the brane \cite{shiromizu}.
\par
In the present letter we investigate spherically symmetric
solutions to Eqs.~(\ref{Deq}) with $D=4$ of the form
\be
ds^2=-N(r)\,dt^2+A(r)\,dr^2+r^2\,d\Omega^2
\label{g}
\ee
with $d\Omega^2=d\theta^2+\sin^2\theta\,d\phi^2$,
which might represent black holes in the brane-world
\cite{chamblin,maartens,kanti,bh,ch}.
First of all, let us recall that the Schwarzschild
four-dimensional metric (\ref{g}) with
\be
N=A^{-1}
\label{AN}
\ee
and $N=1-{2\,M/ r}$ is ruled out as a physical candidate
since its unique propagation in the bulk is a black string
with the central singularity extending all along the extra
dimension and making the AdS horizon singular \cite{chamblin}.
Further, this case is also unstable under linear perturbations
\cite{gregory}.
A few different cases have been recently investigated with
the condition (\ref{AN}) \cite{maartens,kanti}.
We stress that while Eq.~(\ref{AN}) is accidentally verified
in four dimensions, there is no reason for it to hold in this
scenario as well.
It is easy to show that the most general solution satisfying
Eq.~(\ref{AN}) is of the ``Reissner-Nordstr\"om'' (RN) type
\cite{maartens}
\be
N=1-{2\,M\over r}+{Q\over r^2}
\ ,
\label{RN}
\ee
where $Q$ can be interpreted as a ``tidal charge''.
If instead one insists on requiring the Schwarzschild metric
on the brane but with a regular AdS horizon the price to pay is
to have matter in the bulk \cite{kanti}.
\par
We shall here present two new families of analytic solutions
of the form (\ref{g}) on the brane (at $z=0$) obtained by
relaxing the condition (\ref{AN}).
They are determined by fixing alternatively $N$ or $A$ as
in Schwarzschild [so to have the correct $O(1/r)$ behavior
at large $r$] and finding the most general solutions for
the constraints (\ref{Deq}).
These solutions will be expressed in terms of the ADM mass $M$
and the parameterized post-newtonian (PPN) parameter $\beta$
which affects the perihelion shift and Nordtvedt effect
\cite{will}.
The momentum constraints are identically satisfied by
the form (\ref{g}) and the ``Hamiltonian'' constraint
can be written out explicitly in terms of $N$ and $A$ as
\be
&&{1\over 2}\,{N''\over N}
-{1\over 4}\,\left({N'\over N}\right)^2
-{1\over 4}\,{N'\over N}\,{A'\over A}
-{1\over r}\,\left({A'\over A}-{N'\over N}\right)
\nonumber \\
&&
-{1\over r^2}\,(A-1)=0
\ .
\label{H}
\ee
\par\noindent
{\bf Case I.}
By demanding that
\be
N=1-{2\,M\over r}
\ ,
\label{g_tt0}
\ee
the general solution to Eq.~(\ref{H}) is
\be
A={\left(1-{3\,M\over 2\,r}\right)\over
\left(1-{2\,M\over r}\right)\,
\left[1-{M\over 2\,r}\,(4\,\beta-1)\right]}
\ .
\label{g_tt}
\ee
The only other non-vanishing PPN parameter (which
controls the deflection and time delay of light \cite{will})
$\gamma=\beta$ and one knows that $\beta\simeq\gamma\simeq 1$
from solar system measurements \cite{will}.
In particular the combination
$\eta\equiv 4\,\beta-\gamma-3=3\,(\beta-1)$
measures the difference between the inertial mass and the
gravitational mass of a test body.
We also note that the solution (\ref{g_tt}) depends on just one
parameter and for $M\to 0$ one recovers the Minkowski vacuum.
The same is true for the solution (\ref{RN})
provided one defines $Q=2\,M^2\,(\beta-1)$, which yields
\be
N={1\over A}=1-{2\,M\over r}+{2\,M^2\over r^2}\,(\beta-1)
\ .
\label{Q}
\ee
As was noted in \cite{ch,maartens}, for a (point-like)
matter source located on the brane $Q$ should be related
to $M$ through the brane tension, therefore measuring
$\beta$ would give information on the vacuum energy
of the brane-world (or, equivalently, $\Lambda$).
Finally, the metric (\ref{Q}) can be experimentally
distinguished from the case presented here since
the corresponding $\gamma=1$ and does not depend on
$\beta$.
\par\noindent
{\bf Case II.}
Upon setting
\be
A^{-1}=1-{2\,\gamma\,M\over r}
\ ,
\label{g_rr0}
\ee
one obtains
\be
N={1\over \gamma^2}\,
\left(\gamma-1+\sqrt{1-{2\,\gamma\,M\over r}}\right)^2
\ ,
\label{g_rr}
\ee
where now $\gamma=2\,\beta-1$ and this case can be
experimentally distinguished both from  (\ref{Q}) and
case~I above.
Finally, one has $\eta=\gamma-1=2\,(\beta-1)$.
\par
We shall now explore the causal structure of the previous
solutions by expressing the metric elements in terms of
$\eta\sim\beta-1\not=0$ (keeping in mind that $|\eta|\ll 1$
from experimental data \cite{will}).
In particular, we shall show that the sign of $\eta$ is
of great relevance.
\par
\noindent
{\bf Case I.} The metric components are given
by Eq.~(\ref{g_tt0}) and
\be
A=\frac{\left(1-\frac{3M}{2r}\right)}
{\left(1-\frac{2M}{r}\right)\,
\left[1-\frac{3M}{2r}\,(1+\bar\eta)\right]}
\ ,
\label{mepri}
\ee
where we have defined $\bar\eta\equiv (4/9)\,\eta$.
As in Schwarzschild, the event horizon is at
$r=r_h\equiv 2\,M$ and the related Hawking
temperature is
\be
T_H=\frac{\sqrt{1-3\,\bar\eta}}{8\,\pi\,M}
\ .
\ee
\begin{figure}
\centering
\raisebox{4cm}{$ $}
\epsfxsize=3.2in
\epsfbox{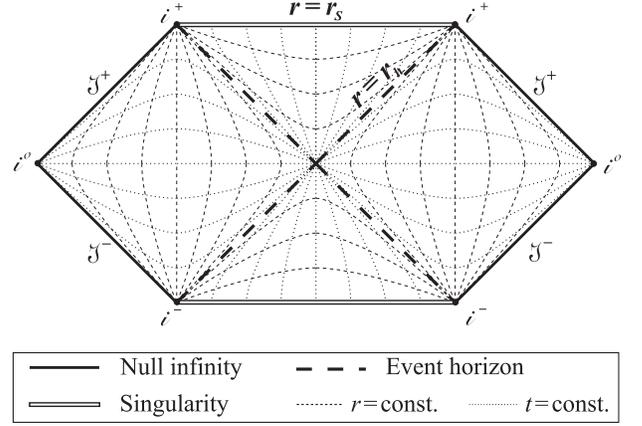}
\hspace{-0.2in}
\raisebox{0.5cm}
{\hspace{6cm} $ $}
\caption{Penrose diagram for case~I and $\eta<0$.}
\label{fig1}
\end{figure}
\noindent
We see that, with respect to the case $\bar\eta=0$,
$T_H$ is either slightly reduced or augmented depending
on the sign of $\bar\eta$.
Inside the event horizon, the line element presents
singularities in the metric for $\bar\eta\not=0$ at
\be
r=\left\{
\begin{array}{l}
\frac{3}{2}\,M\,(1+\bar\eta)\equiv r_0
\\
\\
\frac{3}{2}\,M\equiv r_s
\ .
\end{array}
\right.
\ee
Note that for $\bar\eta>0$ it is $r_s<r_0$, whereas for
$\bar\eta<0$ we have $r_s>r_0$.
Calculation of the curvature invariants
\be
R^2\equiv R_{\alpha\beta}\,R^{\alpha\beta}
\ ,\ \ \
K^2\equiv
R_{\alpha\beta\gamma\delta}\,R^{\alpha\beta\gamma\delta}
\ ,
\ee
shows the presence of a physical singularity at $r_s$,
where $R^2\sim K^2\sim \bar\eta^2/(r-r_s)^4$.
Therefore as $\bar\eta<0$ the space-time ends at $r=r_s$
and the Penrose diagram, similar to that of Schwarzschild,
is represented in Fig.~\ref{fig1}.
\par
\begin{figure}[t]
\centering
\raisebox{4cm}{$ $}
\epsfxsize=3.2in
\epsfbox{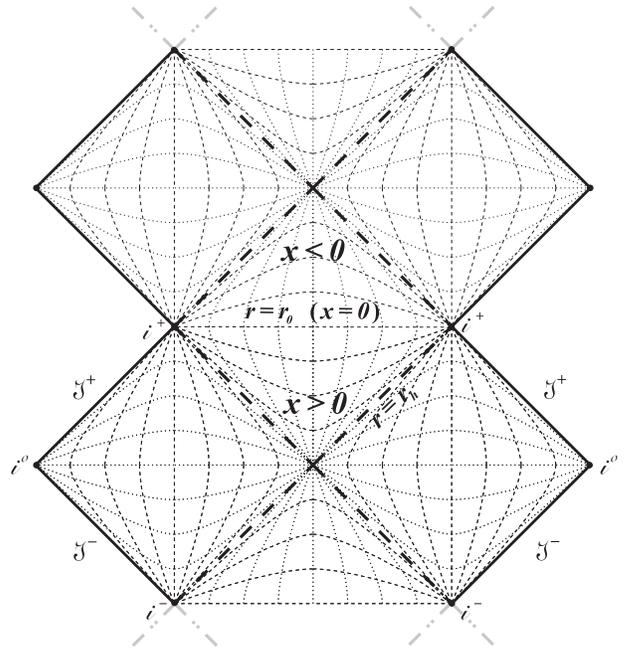}
\hspace{-0.2in}
\raisebox{0.5cm}
{\hspace{6cm} $ $}
\caption{Penrose diagram for case~I and $\eta>0$.
The full diagram can be obtained by repeating the same
structure infinitely many times both in the future and
past.}
\label{fig2}
\end{figure}

The case $\bar\eta>0$ is, in a sense, more interesting and
deserves further investigation.
For $r=r_0$ the squared surface gravity
\be
\kappa^2={M^2\over r^4}\,{r-r_0\over r-r_s}
\ ,
\ee
vanishes (an analogue surface is present in RN in the
region between the two horizons) and the curvature
invariants are regular [they behave as
$\sim 1/(\bar\eta^2\,M^4)$].
Inspection of the equation for the geodesics of energy $E$
and angular momentum $L$,
\be
\left({dr\over d\tau}\right)^2=
{r-r_0\over r-r_s}\left[E^2
-\left({L^2\over r^2}-\epsilon\right)\left(1-{r_h\over r}
\right)\right]
\ ,
\label{geo1}
\ee
shows that the space-like surface $r=r_0$ is a turning point
for all types of curves ($\epsilon=0,\pm 1$ respectively for
null, time-like and space-like geodesics).
A similar phenomenon occurs in RN, where all curves with
$L\not=0$ get reflected at some point
$r=r_f(E,L)<r_-\equiv M-\sqrt{M^2-Q}$.
However, space-like and null curves with $L=0$ are able to
reach the RN time-like ``repulsive'' singularity $r=0$.
In our case no curve is able to enter the region $r<r_0$,
where the signature of the metric is Euclidean.
This makes a substantial difference since the true
singularity at $r_s<r_0$ does not belong to the physical
space.
Furthermore, integration of Eq.~(\ref{geo1}) for $r\sim r_0$
yields the proper time $\tau\sim\sqrt{(r-r_0)\,(r_0-r_s)}$
and continuation of the physical trajectories across
$r=r_0$ can be achieved, e.g., by introducing the coordinate
$x\equiv\sqrt{r-r_0}$ for $r>r_0$ and then going to negative
values of $x$.
In the $(t,x,\theta,\phi)$ coordinate frame, the metric
is given by
\be
&&ds^2=-\left(\frac{x^2+r_0-r_h}{x^2+r_0}\right)\,dt^2
\nonumber \\
&&+\frac{4(x^2+r_0)(x^2 +r_0-r_s)}{x^2+r_0-r_h}\,dx^2
+(x^2 +r_0)^2\,d\Omega^2
\ .
\label{mepri1}
\ee
Since the metric in Eq.~(\ref{mepri1}) is
even under $x\to -x$, both sides of $x=0$ have the same
causal structure.
Indeed, in the region $x<0$ one can introduce a new
radial coordinate $r'$ such that $x=-\sqrt{r'-r_0}$
for which the solution looks exactly like that in
Eqs.~(\ref{g_tt0}) and (\ref{mepri}).
The full Penrose diagram is given in Fig.~\ref{fig2}.
Unlike RN, the space-time is completely regular (the
geometry is that of a traversable wormhole with the minimal
sphere inside the horizon) and continuation beyond the
Cauchy horizon is determined solely by boundary conditions
at asymptotically flat regions.
Regular four-dimensional black holes were constructed in
\cite{frolov} by matching Schwarzschild with de~Sitter
along a space-like surface at $r_j<r_h$, and by
gluing black hole with white hole metrics in \cite{thooft}
as candidate space-times where information is not
lost during the evaporation process.
Contrary to those cases, it is important to stress
that the continuation of our solution across $x=0$ is
smooth, as the extrinsic curvature is continuous
in $x$ (and vanishes at $x=0$).
\par\noindent
{\bf Case II.}
The metric elements can now be written
\be
\begin{array}{l}
N=\left[\frac{\eta
+\sqrt{1-\frac{2\,M}{r}\,(1+\eta)}}{1+\eta}\right]^2
\\
\\
A=\left[1-\frac{2\,M}{r}\,(1+\eta)\right]^{-1}
\ .
\end{array}
\label{mese}
\ee
We shall again distinguish between the two cases $\eta<0$
and $\eta>0$.
\par
For $\eta<0$ the metric is singular at
\be
r=\left\{
\begin{array}{l}
\frac{2\,M}{1-\eta}\equiv r_h
\\
\\
2\,M\,(1+\eta)\equiv r_0
\ ,
\end{array}
\right.
\label{rhh}
\ee
where $r_h>r_0$.
$r_h$ defines the event horizon with formal Hawking
temperature
\be
T_H=\frac{(1-\eta)^2}{8\,\pi\,M}
\ .
\ee
However, the null surface $r=r_h$ is singular since
\be
R^2\sim K^2\sim {\eta^2\over M^3\,\left(
\sqrt{r-r_0}-\sqrt{r_h-r_0}\right)^2}
\ .
\ee
\begin{figure}
\centering
\raisebox{4cm}{$ $}
\epsfxsize=1.6in
\epsfbox{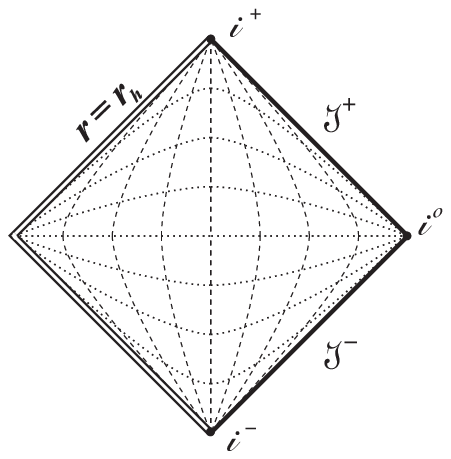}
\epsfxsize=1.6in
\epsfbox{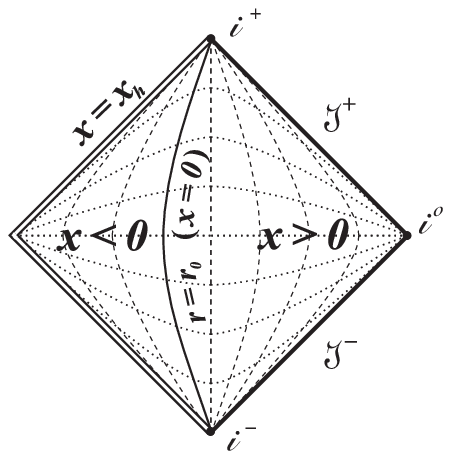}
\hspace{-0.2in}
\raisebox{0.5cm}
{\hspace{6cm} $ $}
\caption{Penrose diagram for case~II with $\eta<0$ (left)
and $\eta>0$ (right).}
\label{fig3}
\end{figure}
\noindent
The corresponding Penrose diagram is represented by the
left diagram in Fig.~\ref{fig3}.
\par
Turning to the case $\eta>0$, we note that the only
singularity in the metric is at $r=r_0$, where all curvature
invariants are regular.
Similarly to the corresponding case~I, the geodesic
equation near $r_0$ is
\be
\left({dr\over d\tau}\right)^2\sim
{r-r_0\over r_0}\left[{(1+\eta)^2\over\eta^2}\,E^2
-{L^2\over r_0^2}+\epsilon\right]
\ ,
\label{geo2}
\ee
and $r=r_0$ is a turning point for all physical curves.
Smooth continuation at $r=r_0$ is achieved by considering
negative values of $x\equiv \sqrt{r-r_0}$.
In terms of the radial coordinate $x$ the line element is
\be
ds^2&=&
-{1\over (1+\eta)^2}\,
\left(\eta+{x\over\sqrt{x^2+r_0}}\right)^2\,dt^2
\nonumber \\
&&+4\,(x^2+r_0)\,dx^2+(x^2 +r_0)^2\,d\Omega^2
\ .
\label{mepri2}
\ee
Unlike the corresponding case~I, however, the two sides of
$x=0$ are not symmetric and the metric for negative $x$
exhibits a singular event horizon at
$x_h\equiv -\sqrt{r_h-r_0}$ [$r_h$ is given in
Eq.~(\ref{rhh}) and
$R^2\sim K^2\sim\eta^2/(M^3\,(x-x_h)^{2})$].
The causal structure is described by the right diagram in
Fig.~\ref{fig3}.
The difference with respect to the left diagram is that the
physical radius $r=x^2+r_0$ reaches a minimum at the
time-like surface $r=r_0$ ($x=0$) and then re-expands when
$x$ turns to negative values.
\par
An interesting aspect from the above analysis is that there
seems to be a correlation between the sign of $\eta$ (a quantity
measured at infinity) and the geometric structure of the
solutions at small $r$ (i.e, of the order of the
Schwarzschild radius for $|\eta|\ll 1$).
For $\eta<0$, a typical trajectory approaching and possibly
entering the horizon is such that the physical radius always
decreases (as in Schwarzschild, i.e., $\eta=0$) and hits the
singularity at $r_s>0$ (larger than $r_s=0$ for $\eta=0$).
Considering positive values of $\eta$, there is always a
turning point at $r=r_0$, as an anti-gravity effect
occurring very close to $r_h$ in regions not yet
experimentally tested. 
A similar feature is present for the metric (\ref{Q}),
\be
N={1\over A}=1-{2\,M\over r}+{\eta\,M^2\over 2\,r^2}
\ .
\label{q}
\ee
For $\eta<0$ there is a single horizon at
$r_h=M\,(1+\sqrt{1-\eta/2})$ with the Hawking temperature
\be
T_H=
{1\over 2\,\pi\,M}\,{\sqrt{1-\eta/2}\over
\left(1+\sqrt{1-\eta/2}\right)^2}
\ ,
\label{TH}
\ee
and a physical space-like singularity at $r_s=0$
(the Penrose diagram is the same as that in Fig.~\ref{fig1}).
If $0<\eta<2$ (RN), there are two horizons at
$r_\pm=M\,(1\pm\sqrt{1-\eta/2})$ [the event horizon
is $r_h=r_+$ and the Hawking temperature is still given
by Eq.~(\ref{TH})] and the time-like singularity at $r_s=0$
becomes ``repulsive''.
It would be interesting to inspect whether the sign of
$\eta$ plays the same role in general.
From a physical point of view, since anti-gravity
effects on the brane are expected for negative brane
vacuum energy $\sigma$ \cite{shiromizu}, we suspect
that (at least for the cases considered here) the sign
of $\eta$ is minus the sign of $\sigma$.
\par
We also note that for finite values of $\eta$ each family
of solutions possesses a zero temperature black hole:
$\bar\eta=1/3$ for case~I, $\eta=1$ for case~II and
the well-known extreme RN $\eta=2$ in Eq.~(\ref{q}).
Contrary to the extreme RN which is singular at $r=0$,
the first two solutions are instead completely regular
and, although the corresponding values of $\eta$ are
ruled out on atrophysical scales, they might be
acceptable candidates as small black holes \cite{bh,ch}.
We will give more details elsewhere.
\par
Let us finally mention that it will be important to
investigate the extension of our solutions into
the bulk.
For Schwarzschild, the singularity at $r=0$ on the brane
extends into the bulk and makes the AdS horizon singular
as well.
However, for $\eta>0$ the solutions with (\ref{g_tt0})
and (\ref{g_tt}) are remarkably free of singularities
and one might hope that the bulk is regular as well.
This study can be attempted either numerically or by
Taylor expanding all five-dimensional metric elements in
powers of the extra coordinate.
The latter method is currently being investigated.
\par\noindent
{\bf Note added}.
After completion of this work we learned that our
case I was also derived in Ref.~\cite{germani} as a
possible metric outside a star on the brane.
In \cite{germani} no attempt was made to study the
causal structure of the space-time.
\par\noindent
{\bf Acknowledgements}.
We wish to thank R.~Balbinot, G.~Cl\'ement, S.~Farese,
B.~Harms, N.~Kaloper and G.~Venturi for useful discussions.
\end{document}